# Comment: A Bragg glass phase in the vortex lattice of a type II superconductor

In a recent *Letter to Nature*, Klein et al.[1] provided strong experimental evidence for the existence of the Bragg glass phase in the vortex lattice of impure type II superconductors. We agree in principle with their interpretation of the neutron-diffraction data of the vortex lattice in favour of the Bragg glass phase. This nice piece of experimental work certainly helps to clarify the existence of a phase, which attracted the attention of many physicists in the past. However, we believe that in Ref.1 the presentation of theoretical work relevant to explain the experiments gives a somewhat biased picture of previous as well as very recent theoretical achievements, and even prevents a possibly more satisfying interpretation of the data.

The stability of three dimensional pure crystals against weak thermal fluctuations is based on the fact that the relative displacement of any two atoms remains finite, even if they are located at arbitrarily distant sites. This property implies the existence of Bragg peaks in the structure factor[2]. In two dimensions, where thermal fluctuations lead to a weak, logarithmic, divergence of the displacement with atom distance, Bragg peaks still exist but exhibit power-law divergences with non-universal exponents. For 3D vortex lattices in impure superconductors, a quite similar logarithmic increase of displacements was predicted by one of us[3] in 1990, stressing the importance of the previously disregarded periodicity of the flux line array. This result was subsequently supported in a more detailed replica theory by Korshunov[4], whose contribution was ignored in Ref. 1. These previous works were confirmed and extended by Giamarchi and Le Doussal[5,6], who also pointed out, in analogy to the situation for pure 2D crystals, the existence of power-law Bragg peaks, thus proposing the vivid name "Bragg glass" for a phase, which was called before "Elastic vortex glass"[7]. The latter name expressed the fact that dislocations were excluded from the vortex array — first by assumption[3,4,5] and later by detailed calculations[8,9,10,11,12]. Taking this into account, the statement of Klein *et al.*[1] about the first prediction of this novel phase in Ref. 6 appears not to be justified. After all, phases are characterized by their properties, not by names.

The authors of Ref. 1 also omit recent work by Emig *et al.*[13,14], which avoids the shortcomings of Flory like theories and completes the analogy with 2D crystals, showing that the Bragg peak exponent $\eta$ is non-universal and depends on the external magnetic field. Although the non-universality is rather weak due to a perturbative expansion of the exponent $\eta$ near the critical dimension of the model, a much more pronounced effect has to be expected in the case of three dimensional lattices. In Refs. 13,14 the existence of *sharp* crossovers between three distinct scaling regimes for the relative vortex displacements could also be clearly identified, providing accurate information about exponents and crossover length scales as, e.g., the positional correlation length $R_a$. These results are, in fact, relevant for testing the consistency of the existence proof for the Bragg glass as given by Klein *et al.*[1]: The height of the Bragg peaks, scaling as $R_a^\eta/a_0 \sim 1/B^\mu$ with lattice spacing $a_0$, as function of the magnetic field $B$ is compared in Fig. 3 of Ref. 1 to theory, assuming a field *independent* Flory exponent of $\mu=3/2$. However, better consistency between theory and experiment could be obtained by considering the field dependence of $\mu=(-1-2\eta+\eta/\zeta)/2$, where the also non-universal[13,14] exponent $\zeta$ describes the displacement growth on length scales below $R_a$.

We think that this Comment could give a more balanced picture of the theoretical developments, which led to the current understanding[15,16] of the Bragg glass phase and offers a better way for the interpretation of the experimental data presented by Klein *et al.*[1]

**S. Bogner, T. Emig, T. Nattermann, S. Scheidl**

Institut für Theoretische Physik, Universität zu Köln, Zülpicherstrasse 77, 50937 Köln, Germany